# A geometry-based relaxation algorithm for equilibrating a trivalent polygonal network in two dimensions and its implications

Kai Xu

*Fisheries College, Jimei University, Xiamen, 361021, China*

kaixu@jmu.edu.cn

ORCiD: http://orcid.org/0000-0002-1341-1525

# A geometry-based relaxation algorithm for equilibrating a trivalent polygonal network in two dimensions and its implications

**Abstract**  The equilibration of a trivalent polygonal network in two dimensions (2D) is a universal phenomenon in nature, but the underlying mathematical mechanism remains unclear. In this study, a relaxation algorithm based on a simple geometrical rule was developed to simulate the equilibration. The proposed algorithm was implemented in Python language. The simulated relaxation changed the polygonal cell of the Voronoi network from an ellipse's inscribed polygon toward the ellipse's maximal inscribed polygon. Meanwhile, the Aboav-Weaire's law, which describes the neighboring relationship between cells, still holds statistically. The succeed of simulation strongly supports the ellipse packing hypothesis that was proposed to explain the dynamic behaviors of a trivalent 2D structure. The simulation results also showed that the edge of large cells tends to be shorter than edges of small cells, and vice versa. In addition, the relaxation increases the area and edge length of large cells, and it decreases the area and edge length of small cells. The pattern of changes in the area of different-edged cells due to relaxation is almost the same as the growth pattern described by the von-Neumann-Mullins law. The results presented in this work can help to understand the mathematical mechanisms of the dynamic behaviors of trivalent 2D structures.

## 1. Introduction

Many natural structures can be simplified as a two-dimensional (2D) network of convex polygons (cells). If every three edges meet at a vertex, the network can be regarded as a trivalent 2D structure, which can be described by three universal laws: Euler's law, Lewis's law, and Aboav-Weaire's law [1]. The latter two empirical laws can be summarized as follows: a small cell tends to neighbor with a small number of larger cells and vice versa. The formulas of these laws indicate deep relationships between local and global structures, and between topology and geometry. The patterns and scales of dynamics of cellular growth are quite different between living and non-living trivalent 2D structures, which indicates that these laws can be achieved and maintained by different physical, chemical, and/or biological mechanisms. Moreover, the Voronoi diagram, which is a type of static structure, is generated on the basis of just one simple geometrical rule, which also obeys the above three laws. Thus, only mathematical mechanisms can ultimately explain all different types of 2D space-filling with convex polygons.

In recent decades, several physical models have been developed to test competing hypotheses and simulate the dynamic behavior of trivalent 2D structures [2, 3]. Some of the models, such as the force-based vertex models, can conveniently trace geometrical and topological changes in all cells during the simulated growth of structures. This is an important advantage over the other models. Even in the absence of a mathematical verification, force-based vertex models have been widely used in many research fields [2-4]. As a core component of vertex models, the relaxation algorithm deals with the problem of equilibrating the network. The natural relaxation can be easily observed by comparing the time-series images of biological 2D structures, especially changes in the angle directly associated with the newborn vertices [5]. Generally, relaxation rounds the cell and concentrates most of the angles about 120°, which must be achieved by isometric growth of the cell wall [5, 6]. However, knowledge is still limited to the dynamic behaviors of cell area, angle, edge number, and edge length of trivalent 2D structures.

Recently, based on the observation that the polygonal cell can be approximately regarded as an ellipse's inscribed polygon (EIP) and tends to form an ellipse's maximal inscribed polygon (EMIP), Xu (2019, 2020) proposed an Ellipse Packing hypothesis to explain Lewis's law and Aboav-Weaire's law and updated these laws' formulas [6, 7]. One of the most significant differences between living and non-living 2D structures is that the cell area of the former is about 0.8 times the EMIP, while that of the latter is very close to the EMIP [6]. This can be attributed to the different time scales of growth processes. At the cellular level, biological-controlled processes take much more time to change cellular geometrical and topological parameters smoothly, and the relaxation can be frequently disturbed by new divisions, but physically or biologically induced chemical processes happen much faster [1, 3, 6, 7]. To test the ellipse packing hypothesis, herein we developed a relaxation algorithm based on only one simple geometrical rule related to the EMIP. This study also analyzed the effects of relaxation on cell area, angle, and edge length of trivalent 2D structures.

## 2. Methods

### 2.1 Relaxation algorithm

In our proposed geometry-based relaxation algorithm, the motion of vertices changes the shape and position of a cell. This is the same as in the classical vertex model. The remarkable feature of our algorithm is that the motion is determined by only one geometrical rule; namely, the difference between the eccentric angles of two neighboring vertices of an $n$-edged EMIP is $2\pi/n$ [8]. When the ratio of long axis to short axis of an ellipse equals to 1, the eccentric angle of a point on the ellipse equals to the rotation angle of the point. Thus, it is assumed that the consequence of vertex motion is that the difference between the rotation angles of two neighboring vertices tends to equal to $2\pi/n$. The step-by-step procedure of the proposed relaxation algorithm can be summarized as follows.

(1) The first step is to find the target positions of vertices. For an $n$-edged polygonal cell, connect the center $o$, which is set as the origin of coordinate system for convenience, with vertices $v_1, v_2, \ldots, v_i$, where $i = 1, 2, \ldots, n$, and $i$ represents the rank in an ascending order of values of angle $\theta_i$ between the positive $x$-axis and line $ov_i$ in the counterclockwise direction, as shown in Figure 1A. The method of least squares is used to find the optimum set of rays $l_1, l_2, \ldots, l_i$ starting from the center $o$, where the angle between two neighboring optimum rays is $2\pi/n$, and $\delta_i$ is the counterclockwise angle between the $i$th optimum ray and the positive $x$-axis. Thus, $\delta_i = \delta_1 + 2\pi(i-1)/n$. When $\delta_1 = (\sum \theta_n - \pi(n-1))/n$, the value of $\sum(\theta_n - \delta_n)^2$ is the minimum. For each vertex, a triangle is formed by three corresponding optimum rays, and the triangle center is the target position of the vertex motion. Distance $D$ denotes the distance between the vertex and the triangle center, as shown in Figure 1B.

(2) Rank all the vertexes except for the vertexes jointed with edges not shared by two cells according to their values of the maximum vertex angle in descending order, and then move the vertexes one-by-one toward their target positions following the order. The motion distance is $\gamma$ times of $D$. Before moving a vertex, check whether the vertex is already inside the triangle zone, whether the vertex motion will increase the sum of squares of the three vertex angles, and whether the motion will result in an angle large than $\pi$; if any of these three is true, do not move the vertex and jump to the next vertex.

(3) Repeat Steps 1 and 2 $P$ times.

*2.2 Ellipse fitting*

To analyze the topological and geometrical dynamics, each polygonal cell is fitted with a circumscribed ellipse. The area ratio of the EMIP to the ellipse is $0.5n\sin(2\pi/n)/\pi$ [8], and the lowest ratio is about 2.4. The classical least squares (CLS) has been commonly used for ellipse fitting and preferred in finding the best-fitted ellipse. However, the main goal of this work is to find the smallest circumscribed ellipse, which generally has a cost on goodness-of-fitting [6, 7]. The CLS can satisfy our goal for most of the cells. However, for cells with less than five vertices and for a few cells deviated far away from the EMIP, the area ratio of the fitted ellipse to cell area can be far larger than 2.4. Hence, the point interpolation is applied to improve ellipse fitting for the cell with the area ratio large than 3.0 based on the CLS method. The cell's vertices are rotated around the cell center by five degrees in both clockwise and counterclockwise directions. Then, an ellipse is fitted based on the coordinates of $n$ vertices of the cell and $2n$ rotated points. If the area ratio is still larger than 3.0, the rotation degrees are increased to eight or ten. If this does not satisfy the requirements, then a new ellipse is fitted based on a combination of the above $3n$ points and $n$ midpoints of edges.

*2.3 Simulation experiment*

The geometry-based relaxation algorithm was implemented in Phyton language. The Voronoi diagram was generated on the basis of only one simple geometric rule, which is as follows: each Voronoi cell is a region containing all points that are closer to its seed (a specific point) than to any other seed. The seed positions of the inner layer of cells of a regular hexagonal Voronoi diagram were randomly disordered with an irregularity of $k$ [7, 9]. To test the effects of relaxation on the trivalent 2D structure with different edge distribution, five types of disordered Voronoi diagrams were generated with five $k$ values of 0.3, 0.4, 0.6, 0.8, and 1, and set as the initial network for simulation experiments. To smoothly equilibrate the polygonal network, this study set $\gamma = 0.01$ and $P = 200$. The peripheral three layers of cells were excluded from the analysis, as shown in Figure 2. At least 500 cells were analyzed to investigate the effects of the relaxation on each type of disordered Voronoi diagram. To calculate the edge distribution and the corresponding second moment $\mu_2$ precisely, more than 35,000 cells were generated for each type of the Voronoi diagram using software R with 'deldir' package.

## 3. Results and discussion

### *3.1 Edge number distribution*

If a 2D Voronoi diagram is generated on the basis of randomly placed seeds or randomly disordered locations of seeds of a regular hexagonal Voronoi diagram, then its cells obey the Euler's, Lewis's, and Aboav-Weaire's laws and exhibit specific distributions of the edge number, interior angle, and area [6, 7, 9-14]. This is one of the most important features of the 2D Voronoi diagram, and it can be used to develop a new method for randomness tests. For instance, in this study, the coordinates of seeds of the Voronoi diagram are generated based on the numbers extracted from decimals of $\pi$, and it is found that the distribution of cell edge numbers is almost the same as that of the Voronoi diagram generated from pseudo-random numbers (Table 1). The similarity between the Voronoi diagram and natural trivalent 2D structures has been firstly noticed by Honda, and then the Voronoi diagram has been frequently used as an initial network in the vertex model [2, 15]. These findings indicate that randomness can be a key to understanding the trivalent 2D tessellation.

*3.2 Ellipse packing*

According to the ellipse packing hypothesis, for cells of living and non-living trivalent 2D structures, different patterns of growth and topological transformations result in different topological and geometrical performances, but all the cells always tend to form EMIPs [6, 7]. Thus, an effective relaxation algorithm should change the EIP network toward the EMIP network without influencing the edge number and neighboring relationship. Based on the area ratio of a cell to the EMIP, the cells of living and non-living trivalent 2D structures were approximately classified as EIP and EMIP, respectively [6, 7]. When irregularity $k$ equaled to 0.6 or 1, the edge distribution of the disordered Voronoi diagram was very close to the natural structures (Table 1). The effects of the relaxation on typical Voronoi diagrams at $k = 0.6$ and $k = 1$ are shown in Figures 2A-B and 2C-D, respectively. This study confirms that the area of the Voronoi cells can be calculated by

$$\text{Area} \approx 0.5nab\sin\left(\frac{2\pi}{n}\right)\left(1 - \frac{\mu_2}{n}\right), \qquad (1)$$

where $a$ and $b$ denote the semi-major axis and semi-minor axis of the fitted ellipse of the $n$-edged cell, respectively, and $\mu_2$ is the second moment of edge numbers [6, 7]. However, the area of relaxed cells is calculated by

$$\text{Area} = 0.5nab\sin\left(\frac{2\pi}{n}\right). \qquad (2)$$

The calculated area ($A_C$) of both Voronoi and relaxed cells was very close to the real area ($A_R$), as shown in Figure 3A. Moreover, although both $a$ and $b$ of each cell were changed by the relaxation, the Aboav-Weaire's law still held statistically. The total edge number ($nm$) of neighbor cells of the $n$-edged cell did not change by the relaxation and can be expressed as

$$nm = n\left(6 - \frac{a}{b}\right) + \frac{6a}{b} + \mu_2, \qquad (3)$$

where $m$ denote the average edge number of neighbor cells of an $n$-edged cell. The calculated $nm$ ($nm_C$) values of the Voronoi and relaxed cells were very close to the real $nm$ ($nm_R$) value (Figure 3B).

These results indicated that the geometry-based relaxation algorithm could be used to successfully shift an EIPs network to an EMIPs network, and given strong support to the hypothesis ellipse packing. However, the basic assumption of hypothesis ellipse packing need to be modified to that the difference between the rotation angles of two neighboring vertices tends to equal to $2\pi/n$. These results raised the following question: When $\mu_2 > 0$, how close are relaxed cells to EMIPs if the relaxation time is long enough? It is likely that, at least from the perspective of dynamics, a 2D plane cannot maintain a tessellation with different-edged EMIPs. After any type of topological transformation, a former EMIP network will not be shifted to a new EMIP network by relaxation because the rotation angle of vertices need to be adjusted and five points determine a conic on 2D geometry.

### *3.3 Effects of relaxation on area and average edge length*

After relaxation, the average cell area was changed by less than 0.4% (datasheet S1). The distributions of the cell area, perimeter, edge length, and interior angle were significantly changed by the relaxation (Figure 4). The area difference ($\Delta A$) between the Voronoi cell ($A_{VC}$) and the corresponding relaxed cell ($A_{RC}$) was calculated by

$$\Delta A = A_{RC} - A_{VC}. \qquad (4)$$

It was found that the areas of cells with more than six edges were increased by the relaxation, while those of cells with fewer than six edges were decreased; also, cells with six edges were slightly influenced (Figure 5A). Such types of responses were insensitive to changes in $\mu_2$ by varying $k$. This is an important fact in exploring the dynamic mechanisms of trivalent 2D structures. The effects of the relaxation on the size of different-edged cells were almost the same as the growth pattern described by the modified von-Neumann-Mullins law [7, 16]. This law reads

$$\frac{dA}{dt} = K(n-m) = K(n-6)\left(1 + \frac{a}{nb}\right) - \frac{K\mu_2}{n}, \qquad (5)$$

where $A$ denotes the cell area, and $K$ relates to the physical and chemical characteristics of a material. For instance, at the irregularity ($k$) values of 0.6 and 1, the area difference ($\Delta A$) caused by relaxation was positively and linearly related to $n - m$, as shown in Figures 5B and 5C. These results suggest that the direct driver of cellular growth of non-living 2D structures is the equilibration of polygonal networks, and $K$ represents the physical and chemical properties determined by the growth (relaxation) rate of different 2D materials, while $\mu_2$ contributes only slightly. Thus, the status of a non-living trivalent 2D structure is determined by the factors that continuously induce the disturbance of a polygonal network and shape the spontaneous equilibration. As for living 2D structures, the relaxation of a network of polygonal cells can be observed on changes in the angle, which is directly associated with newborn vertices generated by cell division [5]. This study suggests that, due to the relaxation, the size of a cell newly generated by cell division tends to be smaller than that of the quiescent cell, which indicates that the cell needs to employ a series of physical, chemical, and biological mechanisms to sustain the conserved area distribution and edge number distribution.

A linear relationship was found both between cell area $A_R$ and edge number $n$ (Figure 6A) and between cell perimeter and $n$ (Figure 6B). The latter is the so-called Desch's law [17]. Moreover, the simulation results showed that the ratio of area to the perimeter was positively and linearly related to $n$ (Figure 6C). Thus, the relationship between cell perimeter and $n$ is more complex than previously believed. Compared with the cell perimeter, the cell's average edge length ($E_{avg}$) showed opposite trends as responses to $n$ and relaxation (Figure 6B&D). These results suggest that the edge length of large cells tend to be shorter than those of small cells, and the edge length of small cells tend to be longer than those of large cells. This edge-length trend can be used to explain the area distribution. Based on Lewis's law, the area of a convex cell increases with the edge number $n$, so the interior angle of small cells are more possibly less than $0.5\pi$ in contrast to those of large cells. The edge-length trend suggests that long edges make small cells not be too small, and short edges make large cells not be too large.

The perimeter difference ($\Delta P$) and the difference of average edge length ($\Delta E_{avg}$) between the Voronoi cell and the corresponding relaxed cell were calculated using the same method as $\Delta A$. Regardless of the value of $k$, the perimeter difference ($\Delta P$) increased with $n$ (Figure 7A). This trend was similar to the relationship between $\Delta A$ and $n$ (Figures 5A&7A). However, the perimeters of six-edged cells significantly decreased after the relaxation, which was quite different compared with the effect of the relaxation on the area. The changes in cell area $A_R$ and perimeter were also positive linearly related to each other; the relationship for $k = 0.6$ and $k = 1$ are presented in Figures 7B-C, respectively. Similar relationships were between the difference of average edge length ($\Delta E_{avg}$) and $n$ (Figure 7D), and between $\Delta A$ and $\Delta E_{avg}$ at $k = 0.6$ (Figure 7E) and $k = 1$ (Figure 7F). The results indicate that the relaxation tends to increase the edge length and area of large cells, and vice versa.

*3.4 Implications for topological dynamics*

According to Lewis's and Aboav-Weaire's laws, small cells tend to neighbor with large cells, and vice versa. Thus, for an edge shared by two neighboring cells, the relaxation will have opposite effects on the area and edge length at two flanks of the edge. Also, the level of this imbalance will increase with the difference in the area or edge number between two neighboring cells (Figure 4–7). Consequently, if an edge is shared by a large cell and a small cell, this edge will be unstable due to the equilibration of the polygonal network. The cell's topological transformations, such as edge loss or gain, neighbor exchange, death, and division, can happen for edges with the highest imbalance level. To sustain the conserved edge distribution (Table 1), the cell division of living trivalent 2D structures must equally divide a large cell and let the smallest neighbor gain an edge and a vertex [18, 19]. The topological transformations require the re-equilibration of polygonal networks. Accordingly, the trivalent 2D structures are self-driven or self-organized by the cycle of relaxation-topological transformation, while the physical and chemical properties of materials and environmental factors are only limiters and inducers rather than drivers.

*3.5 Lower and upper limits of cell's edge number*

Lewis also pointed out a conserved distribution of edge numbers of living trivalent 2D structures [20], which is as follows: the average edge number is always very close to six, the percentage of hexagonal cells is preferred to be the highest, and the edge number is ranged from four to nine. The average edge number of six has already been proven mathematically based on 2D Euler's law [21]. As a master rule, the 2D Euler's law is preserved under all types of topological transformations. The ranges of edge numbers show a significant difference between non-living and living trivalent 2D structures, as shown in Table 1 [5-7, 9-11, 14]. However, the mechanisms behind the conserved ranges of edge numbers still remain unclear, and this work aims to explain these mechanisms on the basis of the improved Lewis's law [6, 7].

For living trivalent 2D structures, the cell's area, cell's edge number, and cell number are mainly determined by the equal-sized mitosis division that prefers to transect at the midpoint of edges [5-7]. Consequently, the average edge number of two daughters of an $n$-edged cell is $(0.5n + 2)$. If the edge number of a mother cell is three, then one of its daughters contains four edges, which means that according to Lewis's law, the area of that daughter tends to be much larger than that of the three-edged mother. However, this is not consistent with the fact that the area of a daughter cell must be much smaller than that of its mother cell. Thus, the smallest edge number of cells of living trivalent 2D structures is four. As for the soap froth and other non-living 2D structures, the growth in cells is mediated by the area (mass) transfer between neighboring cells and can be described by the von-Neumann-Mullins law [3, 7, 16]; then the smallest edge number can go down to three, which is also the smallest edge number of a polygon. The difference in the smallest edge number of living and non-living cells can be attributed to different topological transformations; namely, the former mainly or solely employs division, and cells of the latter never or rarely divide. The cell number of living structures can be exponentially increased by cell division when supply with enough energy and sources, but the cell number of non-living structures will be decreased by the combination of neighbor exchange (T1 process) and disappearance of a triangle (T2 process) [1-3]. Thus, the topological transformations also determine whether a polygonal network can turn back to its previous topological and geometrical status. To be more precise, the living trivalent 2D structures cannot turn back to their previous topological and geometrical status because the structures and their cells cannot suffer a large loss of mass, while the non-living structures can.

Assume that in Eqs. (1) and (2) both $a$ and $b$ are constant; then the increase in the cell area decreases with the edge number, and the increases of living cells drops much faster than those of non-living cells. Based on the previous studies [6, 7] and the results presented in this study, for living trivalent 2D structures, when the edge number increases from eight to nine, the changes in the area of the fitted ellipse are less than 2%, and the cell area increases by less than 3.5% (assume $\mu_2 = 0.77$, Table 1) due to the increase in the edge number. Thus, the overall increase in the cell area is less than 7%. As for non-living cells, the cell area increases by 0.7% when the edge number increases from 12 to 13. Meanwhile, even the area of the fitted ellipse increases by 10%, the overall increase in cell area will be only 7%. Further increase in the edge number will not significantly affect the cell area. Therefore, the modified Lewis's law [6, 7] can explain why the upper limits of the cell's edge number are about nine and 13 for living and non-living trivalent 2D structures, respectively. Although the topological transformations of 2D amorphous $SiO_2$ are the same as those of many other non-living trivalent 2D structures, its edge numbers range from four to nine (Table 1), which can be attributed to the additional restriction on the bond length and angle [12].

## 4. Conclusion

In this paper, a geometry-based relaxation algorithm that can improve the classic vertex models and help to understand the physical and mathematical mechanisms of the relaxation is proposed. The proposed algorithm can successfully simulate the transition of a cell shape from the EIP toward the EMIP, thus giving strong support to the ellipse packing hypothesis. Based on the results, the following conclusion can be drawn: edges of large cells tend to be shorter than edges of small cells, while edges of small cells tend to be longer than those of large cells. The simulation results show that the relaxation increases the area and edge length of the large cells while decreasing those of the small cells. The obtained results suggest that the growth patterns and topological transformations should be driven by the relaxation, which equilibrates the trivalent polygonal network. This study shed light on the physical and mathematical mechanisms of topological and geometrical dynamics of trivalent 2D structures.

**Acknowledgements**

The author thanks Mr. Guowei Shi and Mr. Jun Song for their technical supports on implementation of relaxation algorithm. Many thanks to the support from my family. We thank LetPub (www.letpub.com) for its linguistic assistance during the preparation of this manuscript.

**Declaration of interest statement**

No potential conflict of interest was reported by the author.

**Table 1** Percentage of the $n$-edged cells of seven types of trivalent 2D structures.

| 2D structures | | Percentage of $n$-edged cell | | | | | | | | | | | Cell number |
|---|---|---|---|---|---|---|---|---|---|---|---|---|---|
| | | 3 | 4 | 5 | 6 | 7 | 8 | 9 | 10 | 11 | 12 | 13 | |
| Pseudo-random-seeded Voronoi | | 1.09 | 10.89 | 25.92 | 29.39 | 19.93 | 9.06 | 2.81 | 0.75 | 0.14 | 0.03 | 0.004 | 47786 |
| Voronoi based on decimals of $\pi$ | | 1.12 | 10.80 | 25.98 | 29.45 | 19.94 | 8.83 | 2.92 | 0.77 | 0.15 | 0.03 | | 47772 |
| Disordered Voronoi | $k = 0$ | | | | 100 | | | | | | | | $\infty$ |
| | $k = 0.3$ | | | 0.90 | 98.21 | 0.89 | | | | | | | 37933 |
| | $k = 0.4$ | | 0.11 | 7.74 | 84.44 | 7.52 | 0.20 | | | | | | 37969 |
| | $k = 0.6$ | 0.11 | 3.23 | 24.12 | 47.13 | 20.75 | 4.15 | 0.47 | 0.05 | 0.003 | | | 38049 |
| | $k = 0.8$ | 0.43 | 7.23 | 26.30 | 35.65 | 21.17 | 7.37 | 1.60 | 0.22 | 0.03 | 0.003 | | 38067 |
| | $k = 1$ | 0.78 | 9.00 | 26.33 | 31.82 | 21.08 | 8.32 | 2.20 | 0.39 | 0.07 | 0.005 | | 38141 |
| Soap froth [11, 14] | | 1.0 | 9.1 | 31.4 | 30.5 | 17.0 | 6.9 | 3.3 | 0.8 | 0.5 | 0.08 | 0.03 | no data |
| Amorphous $SiO_2$ [12] | | | 3.79 | 27.45 | 44.48 | 16.09 | 7.57 | 0.63 | | | | | 317 |
| *Pyropia* [5] | | | 3.96 | 30.44 | 46.84 | 16.56 | 2.34 | 0.09 | | | | | 2253 |
| *Drosophila* [13] | | | 2.95 | 27.90 | 45.72 | 20.12 | 3.18 | 0.14 | | | | | 2172 |

**Figure 1** Diagram of the geometry-based relaxation algorithm.

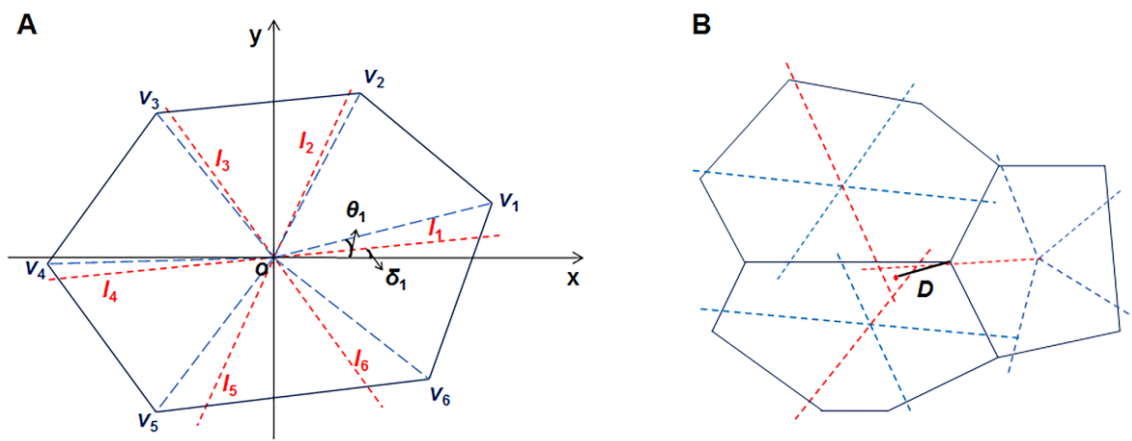

**Figure 2** The Voronoi and the corresponding relaxed networks. Images of two typical disordered Voronoi diagrams and the corresponding relaxed cellular networks at the irregularity of $k = 0.6$ (A),(B) and $k = 1$ (C),(D). The blue dashed line shows the fitted circumscribed ellipse for each cell. Only the cells with their center marked by a black dot were used in the analysis.

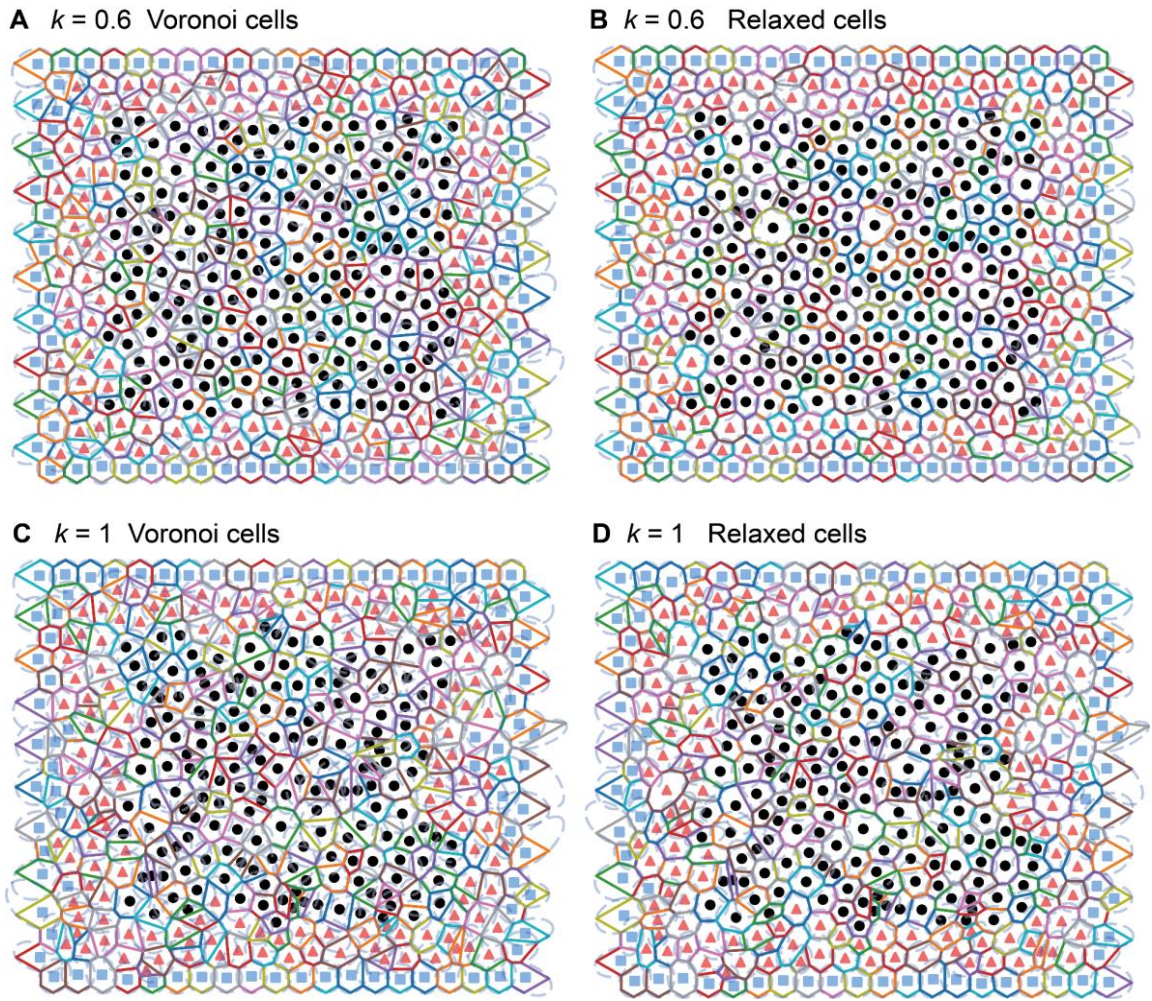

**Figure 3** Effects of the relaxation on the cell area and neighboring relationship. (A) Ratio of calculated area ($A_C$) to real area ($A_R$) of the Voronoi and relaxed cells. (B) Ratio of calculated $nm$ ($nm_C$) to real $nm$ ($nm_R$). $k$ denotes the irregularity of the disordered Voronoi diagram.

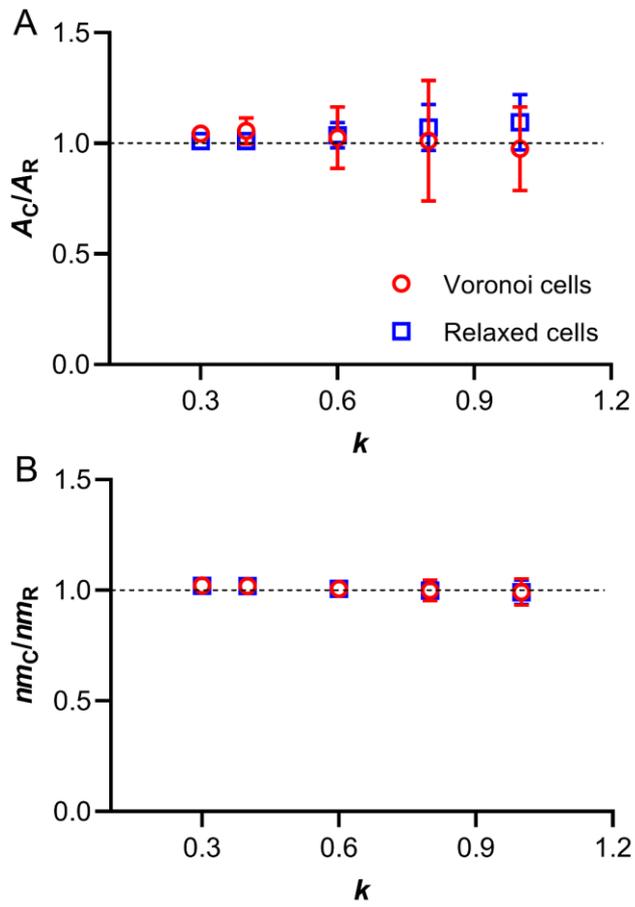

**Figure 4** Effects of the relaxation on the distributions of geometrical parameters. Cell area (A), perimeter (B), edge length (C), and interior angle (D) of a typical disordered Voronoi diagram at the irregularity of $k = 1$ and the corresponding relaxed network.

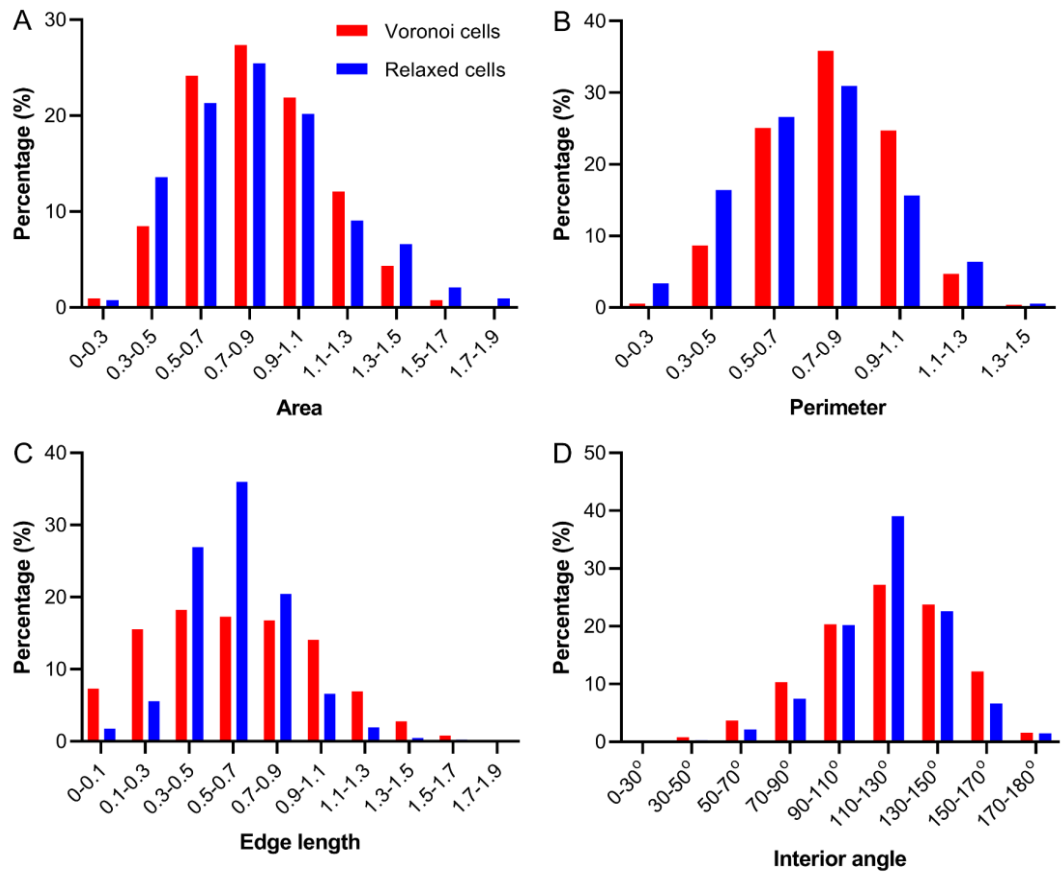

**Figure 5** Effects of the relaxation on the cell area. (A) Relationships between the area change $\Delta A$ and edge number $n$, and (B) between $\Delta A$ and $n - m$ at the irregularity values of $k = 0.6$ and (C) $k = 1$. $m$ denotes the average edge number of neighbors.

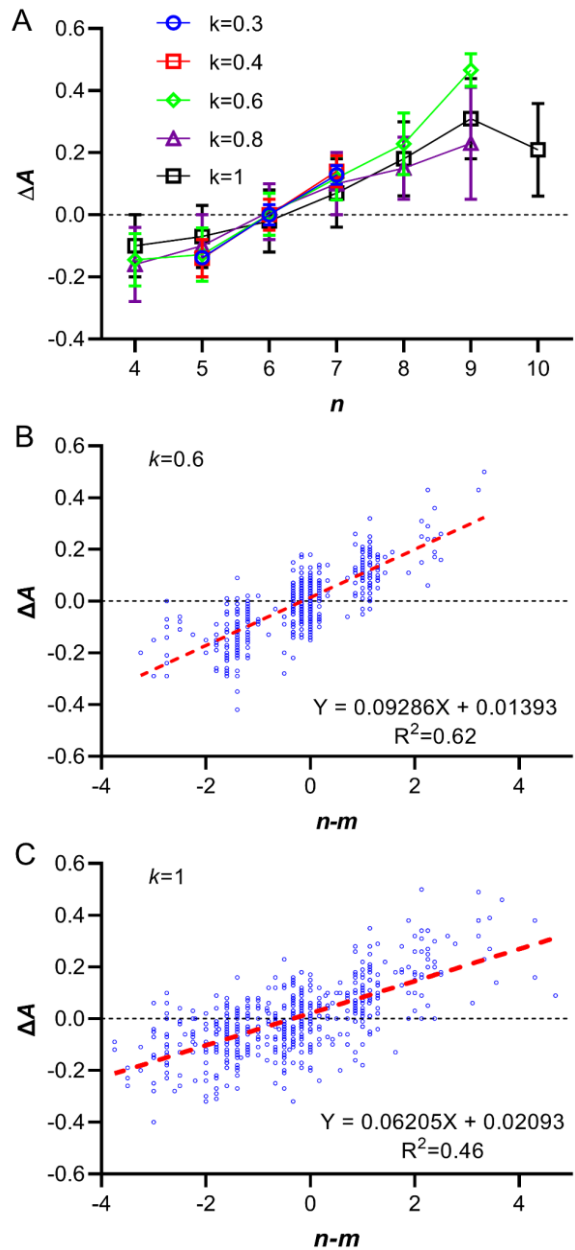

**Figure 6** The relationships between the cell area $A_R$, perimeter, ratio of area/perimeter, and cell's average edge length $E_{avg}$ and edge number $n$.

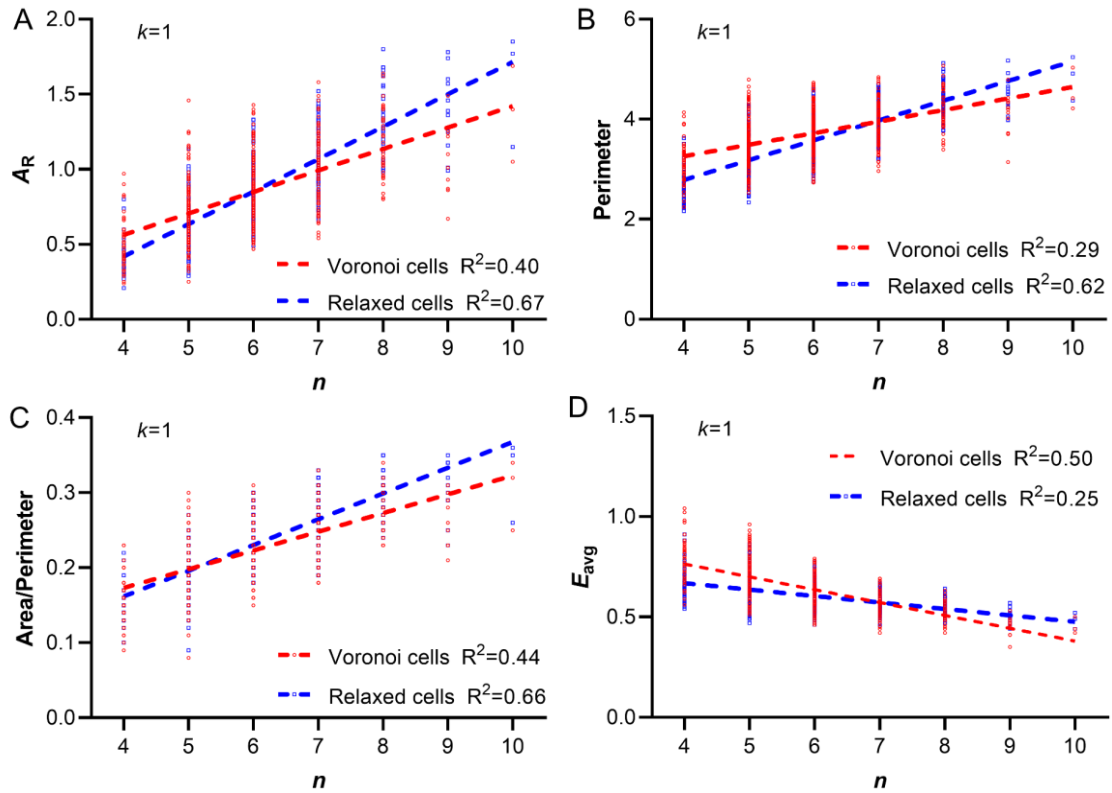

**Figure 7** Effects of the relaxation on the perimeter and a cell's average edge length. Relationships between the perimeter change $\Delta P$ and edge number $n$ (A) and between $\Delta P$ and changes in area $\Delta A$ at the irregularity values of $k = 0.6$ (B) and $k = 1$ (C). Relationships between changes in the cell's average edge length $\Delta E_{avg}$ and $n$ (D), and between $\Delta E_{avg}$ and $\Delta A$ at the irregularity values of $k = 0.6$ (E) and $k = 1$ (F).

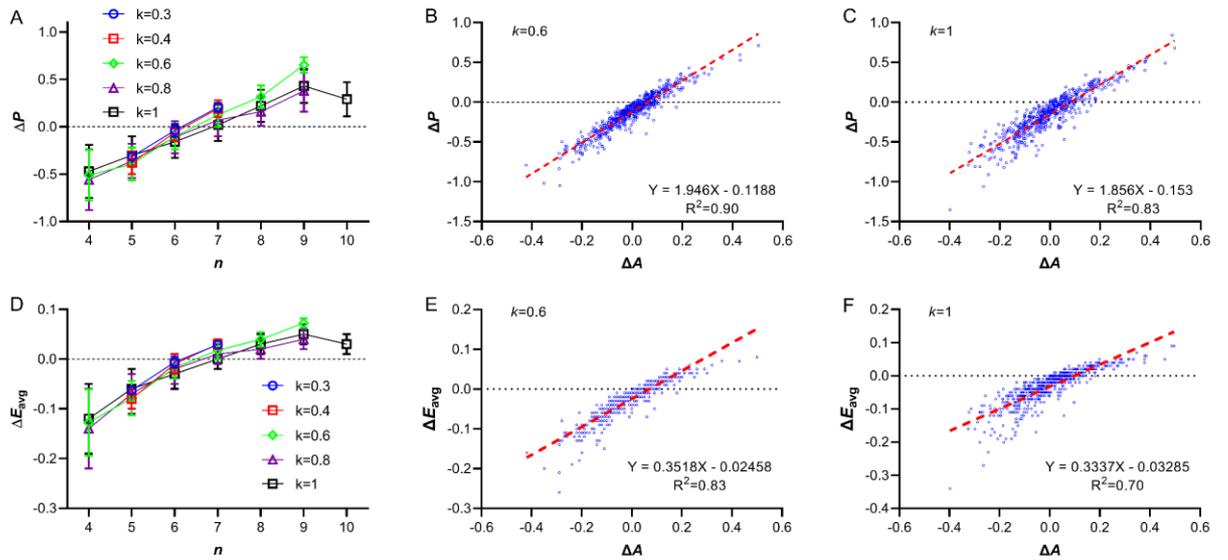